# Robust Learning Based Condition Diagnosis Method for Distribution Network Switchgear

Wenxi Zhang[1], Zhe Li[1], Weixi Li[2], Weisi Ma[2], Xinyi Chen[1], Sizhe Li[3]

(1.Henan Polytechnic University, Jiaozuo, Henan, China, 454000, 2.Jiaozuo Power Supply Bureau, Jiaozuo, Henan, China, 454000, 3.Nanjing University Of Finance & Economics, Nanjing, Jiangsu, China, 210000)

**Abstract:** This paper introduces a robust, learning-based method for diagnosing the state of distribution network switchgear, which is crucial for maintaining the power quality for end users. Traditional diagnostic models often rely heavily on expert knowledge and lack robustness. To address this, our method incorporates an expanded feature vector that includes environmental data, temperature readings, switch position, motor operation, insulation conditions, and local discharge information. We tackle the issue of high dimensionality through feature mapping. The method introduces a decision radius to categorize unlabeled samples and updates the model parameters using a combination of supervised and unsupervised loss, along with a consistency regularization function. This approach ensures robust learning even with a limited number of labeled samples. Comparative analysis demonstrates that this method significantly outperforms existing models in both accuracy and robustness.
**Keywords:** distribution network switchgear; state diagnosis; feature mapping; robust learning; decision radius

## 1 Introduction

In the distribution network operation, 10kV switchgear, ring network cabinet and other distribution network switchgear are extremely important equipment, widely used in the power distribution network system, its reliability and stability is the basis for the effective operation of the power grid, but also to ensure the key to power quality. Nowadays, with the goal of "carbon peak, carbon neutral" and the concept of Internet of Things (IoT) in electric power, many research institutions at home and abroad have carried out engineering practice and technical research to realize the technical requirements of equipment state awareness and multi-source digital fusion, for example, the literature [1-2] has carried out an in-depth study on the state assessment and intelligence of power transmission and distribution equipment. Literature [3] summarized and analyzed the main types of faults of distribution network switchgear as temperature rise faults, refusal to open and close faults and insulation faults. Literature [4-6] analyzed and concluded that temperature monitoring is currently a relatively simple and effective means of intelligence, and this conclusion is based on three-dimensional numerical simplification to get the model of various types of temperature measurement technology. Literature [7-10] quantitatively analyzed the breaking and closing parameters and fault parameters of circuit breakers through online monitoring techniques. Literature [11-16] attempted to analyze the insulation status of the equipment by studying partial discharge detection means, but could not accurately analyze the insulation status of the equipment due to insulation fault mechanisms, fault diagnosis algorithms, and measurements. Literature [17-19] have made many attempts to study the intelligence of distribution switchgear, and as far as the current situation is concerned, the lower degree of intelligence is difficult to meet the requirements of intelligent operation and maintenance in the case of shortage of operation and maintenance personnel. Therefore, the introduction and application of intelligent algorithms in the power Internet of Things to monitor and diagnose the fault status of distribution network switches, improve the intelligent operation and maintenance level of distribution network switchgear, solve the problems of operation and maintenance personnel, and ultimately realize differentiated operation and maintenance has become a current research hotspot.

At present, the diagnostic algorithms of distribution network switchgear are divided into two categories: traditional algorithms and deep learning algorithms. Traditional algorithms are more inclined to rely on manual experience, so the results produced by their algorithms have a certain degree of subjectivity, their robustness is poor, and the accuracy rate is low. For example, literature [20] based on adaptive fuzzy algorithm, the locally generated UV pulse signal and equipment temperature and as input information, in order to detect and judge the cable defects in the switchgear cabinet, this algorithm in the inference rules and affiliation relies on the subjective experience of the expert; Literature [21] proposed a diagnostic algorithm based on the clustering of the FCM method and the correlation analysis, the algorithm in the evaluation factors in the weights are also the subjective judgment of experience. Literature [22] applies models such as deep neural networks to diagnose fault types through autonomous learning by using multiple parameter data as a training set. Parameters include: temperature, electrical parameters, operating status and discharge conditions, etc. However, the large demand for data is a characteristic of the deep learning algorithm class models, so there are some difficulties in practical application.

Based on the above research content, this paper gives full consideration to a variety of factors, on the basis of the distribution switchgear digital twin technology system, the use of multi-source data to build the expansion feature vector of the distribution switchgear, and the use of multi-layer perceptual machine for feature mapping; the introduction of robust learning technology to solve the problem of insufficient generalization ability of traditional methods and large data demand of deep learning algorithms, so as to build a model to diagnose the distribution network switchgear Failure. The main idea of the method is to use labeled data to initially obtain the fault sample center, and then use the unlabeled samples to correct the estimation results of the model. Through the analysis of examples, it is proved that the method proposed in this paper can effectively improve the efficiency of sample use and alleviate the problems of model overfitting, thus significantly improving the fault diagnosis accuracy compared with the existing model.

## 2 Distribution network switchgear digital twin system

The distribution network switchgear digital twin system [23] consists of five parts, as shown in Fig. 1. Among them, the physical equipment layer is the physical object of the digital twin, which has a wide range of total classes and can be used as the information source of the diagnostic model, and the data has the characteristics of nonlinearity, randomness, multidimensionality, and time-varying nature. The physical device layer can provide real-time data and time-recorded report information for the data-aware layer, and can also receive feedback commands from the digital system. The sensing entity of the data perception layer is the signal acquisition and measurement and control device of the equipment, and these sensing entities include the protection measurement and control device on the secondary equipment of the electric power, the sensors on the switchgear, and the signal acquisition device, etc. The role of the sensing entity is to collect the data from the physical equipment. The role of the sensing entities is to collect various dynamic data of the equipment from the physical equipment layer and upload the data to the data center - intelligent station room. The data transmission layer is the link between the end devices and the master and sub-stations. The data transmission layer has an efficient network transmission and data storage system, thanks to the use of network technology, which allows real-time data and information on the operation of the power equipment - data transmission units to be transmitted from the physical devices to the digital twin.

The digital twin can gradually correct the results based on the online real-time data of the equipment on the basis of knowledge and data fusion-driven fault sets and comment sets based on intelligent algorithmic models. The corrected results can reasonably evaluate and predict the state and faults of distribution switchgear, which can more accurately reflect the characteristics of the physical entity, and upload the state evaluation and prediction results to the application platform for early warning of equipment faults, assisting operation and maintenance personnel in managing distribution network switchgear.

The specific deployment method is as follows: in accordance with the principle of primary and secondary integration, and according to the application characteristics of distribution network switchgear, devices such as busbar, contact, outgoing cable contact environmental temperature and humidity sensors, angular displacement sensors, gas density relays and sensors, breaking and closing coil current sensors, local discharge sensors, high-definition cameras, and other intelligent perception terminals are deployed on the switchgear body. Intelligent sensing terminals are installed in the instrumentation room of switchgear, and through hard connection or wireless way to realize the access of sensing devices in a cabinet and in situ analysis, real-time monitoring of switchgear status and in situ linkage.

In addition, the equipment configuration scheme should take into account the user's customized needs, types of distribution network switchgear and application scenarios, and realize differentiated and flexible configuration through the sensing terminal. For example, for application scenarios with general intelligence requirements, only a single temperature measurement device can be equipped, while for application scenarios with higher intelligence requirements, various types of sensing devices can be configured according to demand. Even if only configure the temperature measurement equipment should also be flexible according to the application scenario, such as ring network cabinet can not monitor the dynamic and static contact temperature, while the switchgear according to the need to choose different points of temperature measurement of the differentiated configuration scheme. The hardware and software architecture of the sensing terminal adopts modular design, which can be flexibly combined according to the needs to adapt to the actual needs of different scenarios.

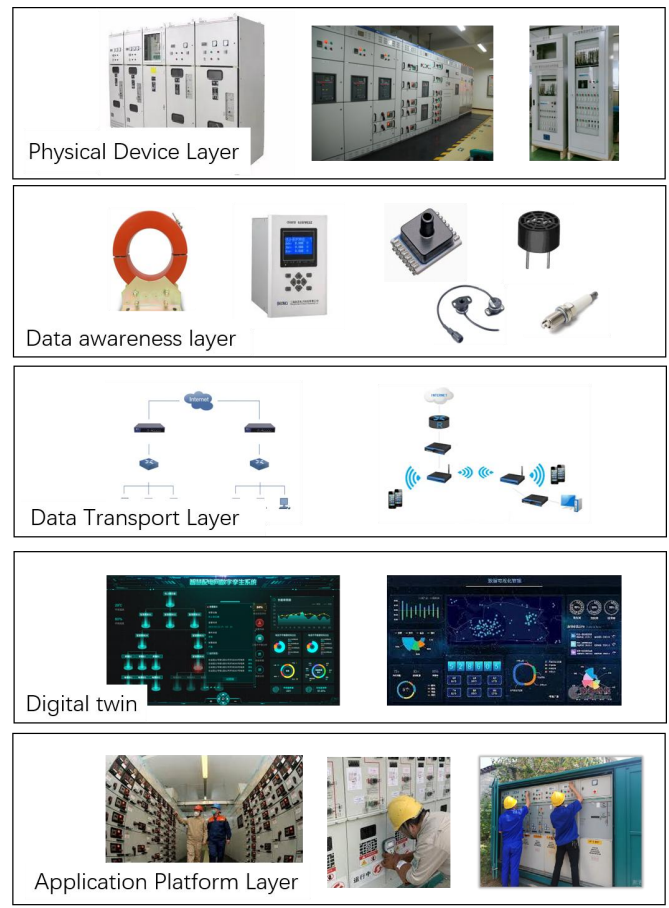

**Fig.1 Digital twin system of distribution switch equipment**

## 3 Distribution network switchgear feature set construction

In order to diagnose the state of distribution network switchgear, it is necessary to clarify the input and output of the model, i.e., the construction of the feature set and the assessment of the state of distribution network switchgear. In the feature set construction part, considering the actual situation and engineering experience, we propose to expand the feature vectors containing multi-source data, in which the position determination of the opening and closing gates adopts the image recognition model mask RCNN, and at the same time, taking into account the computational efficiency and prediction accuracy of sparse coding in the conditions of limited samples, we introduce the feature mapping based on the multilayer perceptual machine. Based on the above input and output data, in order to overcome the problems such as insufficient generalization ability of traditional machine learning under limited multi-source data, robust learning is introduced, the core idea of which is to use unlabeled data to update the model parameters, so as to improve the quality of model parameter estimation, and the method contains four parts, namely, the paradigm learning module, the feature mapping module, the decision radius function and the loss function.

### 3.1 Expanded feature vectors based on multi-source data

As can be seen from the previous summary and analysis, the two most important parts of distribution network switchgear intelligence are temperature monitoring and fault analysis. Thus, it is necessary to obtain the contact temperature measurement of the distribution network switch, the breaking and closing coil current waveform and formation parameters, the equipment state position and motor current and other parameters and conduct comprehensive analysis. In addition, the temperature and power of the outlet cable contact, dynamic and static contacts and other connections of the switching equipment, the contact state, and the aging state are also important factors to be considered.

In addition, switch position visualization and real-time monitoring of current are important factors for the realization of intelligent equipment. It is necessary to accurately observe whether the circuit breaker and grounding knife gate are in place and confirm their status. Real-time monitoring of the motor characteristic current of the switchgear of the energy storage motor, drive motor, etc. can effectively determine whether the motor is in normal working condition. In this paper, the following expansion feature vectors are constructed:

$$\boldsymbol{x} = [r, q, p, l, t, c, f, m]^T \quad (1)$$

Where $\boldsymbol{x}$ is the expanded eigenvector and $r \in \mathbb{R}^3$ is the three-phase outlet temperature.Degree; $q \in \mathbb{R}^6$ is the opening and closing speed, time and stroke respectively; $p \in \mathbb{R}^2$ is the actual opening and closing state of the circuit breaker and the ground knife during opening and closing, with the exception of 0 and the normal of 1; $t \in \mathbb{R}^2$ is the ambient temperature and humidity respectively; $c \in \mathbb{R}^3$ is the ultrasonic value at different positions; $l \in \mathbb{R}^6$ is the running time and peak current of the energy storage motor and the driving motor respectively; $f \in \mathbb{R}$ is insulation resistance; $m \in \mathbb{R}$ is the equipment load rate. The main sources of the above characteristics are temperature, partial discharge, travel, switching current, video monitoring and other sensing equipment. The preprocessing of the original data is as follows: wavelet denoising and removing interference values, and normalizing.

### 3.2 Equipment location judgment based on Mask RCNN network

The preprocessing of the data can initially determine the existence of faults, but the number of switches in the distribution network is large, and it is still necessary to further determine the location of faulty equipment. Through the camera and other auxiliary devices combined with the Mask RCNN network [24] recognition model can achieve the precise location of faulty equipment location, compared with other models, the Mask RCNN network has the ability of target recognition and semantic segmentation, a clear framework, and a strong generalization ability, so it is widely used. The structure of this network is shown in Fig. 2. The process of image device location state recognition based on Mask RCNN network is as follows: firstly, pictures of circuit breakers in various locations and states are collected and labeled as training samples, and the various locations and states of the circuit breakers include: the split position, the intermediate state, the closing position picture, and the ground switch split position as well as the closing position. Secondly, the learned model automatically recognizes the actual position of the switching device in the monitoring video and outputs the results for use by other modules.

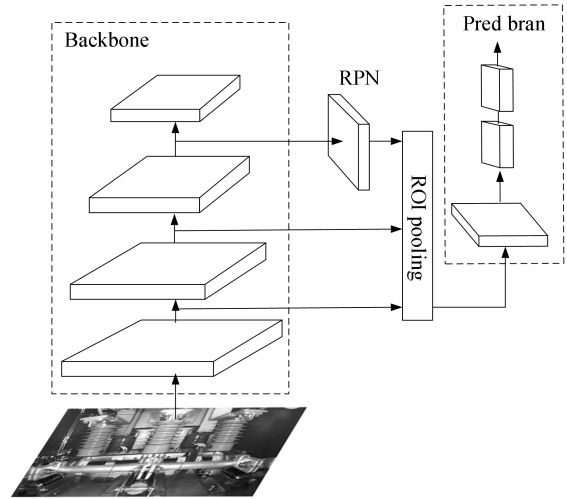

Fig. 2   Image recognition network architecture

### 3.3 Feature mapping based on multilayer perceptron

The essence of the extended feature vector is a high-dimensional random vector, where different dimensions represent the categories in which different feature quantities are located. Under the condition of limited samples, the direct use of the above sparse coding method cannot guarantee the operational efficiency and prediction accuracy [25], so this paper utilizes Multilayer Perceptron (MLP) to transform the original high-dimensional vectors into low-dimensional continuous features, namely

$$\boldsymbol{v} = f_{\text{MLP}}(\boldsymbol{x}) \quad (2)$$

Where $\boldsymbol{v}$ is the mapped feature vector, $f_{\text{MLP}}$ is the multi-layer perceptron model, and $\boldsymbol{x}$ is the original feature vector. That is to say, for the condition diagnosis of distribution network switchgear, the bottom input of the model is the extended feature vector, which passes through the multi-layer perceptron model with $L$ layers and finally outputs the mapped feature vector.

### 3.4 Distribution switchgear state evaluation

The typical state information of distribution switchgear can be divided into seven situations:

1) Normal state.

2) Operating mechanism jamming: when the operating age is long, the circuit breaker mechanism is jammed, resulting in prolonged switch opening and closing time, and in serious cases, problems such as inability to store energy and inability to open and close the gate.

(3) Insulation failure: vacuum tube leakage, reduced sealing, insulation aging triggered by internal breakdown faults; regional air humidity, resulting in operating mechanism corrosion, arc extinguishing chamber and insulating tie rod moisture and other problems.

(4) Empty-closing phenomenon: the resistance of internal moving parts increases under low temperature, and the elasticity of the closing spring of the operating mechanism also decreases, resulting in the mechanism being unable to overcome the load generated by the closing movement of the circuit breaker, and the phenomenon of empty closing occurs.

(5) Mechanical failure: some products have poor anti-vibration performance, in the opening and closing operation, the operating mechanism generates a large vibration force, due to the role of vibration force, so that part of the mechanism connecting rod, connecting piece of the screws and pins between the screws and pins loosened, fall off.

(6) Accidental tripping: Since the external anti-tripping mechanical blocking device function is not configured, there is a risk of sudden tripping of the circuit breaker during the bypass lap lead process. The use of integrated disconnect switches, but the disconnecting knife due to corrosion of the coastal environment, resulting in the presence of oxide layer on the surface of the circuit resistance increases, resulting in abnormal heat.

(7) Problems with secondary equipment: When the FTU is in operation, the unused secondary equipment ports are exposed in the external environment, and the ports are corroded obviously due to the influence of filth and condensation. In high temperature, low temperature, large night temperature difference, and extreme weather, there are problems such as sluggish FTU operation, unstable protection action time, communication dropout, circuit board damage and battery failure.

## 4 Robust Learning-based Condition Diagnosis of Switchgear in Distribution Networks

The core idea of distribution network switchgear state diagnosis is to judge the equipment state based on multi-source sensing data, which is essentially a clustering problem. For the classification problem, the core idea is that the sample features of the same category have similarity. The usual practice is to select the center of each category and calculate the similarity of unknown samples from each center to judge their categories. However, the above method has obvious limitations, and when the training samples are small, the performance of the model decreases significantly. In order to improve the robustness of the model, this paper introduces consistency regularization and decision radius function to process the unlabeled data [26], and the processed data is used for the update of category centers.

The network structure is shown in Fig. 3, in which the main parts include: paradigm learning module, feature mapping module, decision radius function and loss function. Among them, the feature mapping module adopts the multilayer perceptron model described in Section 3.3, while the decision radius function and loss function are the improved parts. The model training process is as follows: firstly, the paradigm learning module generates the corresponding sample set, which contains three parts: training set, test set and unlabeled data. After the data are mapped, the initial category center is calculated. The decision radius function is utilized to predict the categories of the unlabeled data. The above predicted categories will be used to compute the unsupervised loss function. The final loss function contains both supervised and unsupervised components. The supervised loss is used to update the decision radius function, and the unsupervised loss and supervised loss together are used to update the feature mapping module. The specific calculation formula is as follows:

For the training samples in the training set, the category center can be expressed as:

$$c_k = \frac{\sum_m v_m \mathbf{1}_k(y_m)}{\sum_m \mathbf{1}_k(y_m)} \quad (3)$$

Where: $c_k$ represents the center point corresponding to the working state category $k$, $v_m$ represents the $m$-th feature vector, and $\mathbf{1}_k$ is the discriminant function corresponding to the category $k$, that is, $\mathbf{1}_k(y_m) = \mathbf{1}[y_m = k]$.

On this basis, the normalized distance from the $n$ th unlabeled sample to the $k$-th category center is:

$$\tilde{d}_{n,k} = \frac{d_{n,k}}{\frac{1}{M}\sum_{n'} d_{n',k}} = \frac{\|v_n - c_k\|_2^2}{\frac{1}{M}\sum_{n'} \|v_{n'} - c_k\|_2^2} \quad (4)$$

Where: $\tilde{d}_{n,k}$ is the normalized distance, $d_{n,k}$ is the original distance, and $M$ is the number of unlabeled samples.

Based on the statistical data of the above distance information, the formula for calculating the decision radius based on the multilayer perceptron is:

$$r_k = f_{\text{MLP}}(f_{\text{stat}}(\tilde{d}_{n,k})) \quad (5)$$

Where: $r_k$ is the decision radius corresponding to the $k$-th category, $f_{\text{stat}}$ is the statistical data calculation function, including maximum, mean, variance, skewness and kurtosis, and $f_{\text{MLP}}$ is a multi-layer perceptron model.

The decision radius can be used to predict the categories of unlabeled data, that is, for samples falling within the radius $\tilde{d}_{n,k} < r_k$, the probability that they

belong to different categories is:

$$\tilde{y}_n = \operatorname{argmax}_k P(k | v_n, \{c_k\})$$
$$= \operatorname{argmax}_k \frac{\exp(-\|v_n - c_k\|_2^2)}{\sum_{k'} \exp(-\|v_n - c_{k'}\|_2^2)} \quad (6)$$

Where: $\tilde{y}_n$ is the prediction category and $P$ is the conditional probability.

The category center can be updated again using the above prediction categories:

$$c'_k = \frac{\sum_m v_m \mathbf{1}_k(y_m) + \sum_n v_n \mathbf{1}_k(\tilde{y}_n)}{\sum_m \mathbf{1}_k(y_m) + \sum_n \mathbf{1}_k(\tilde{y}_n)} \quad (7)$$

Where: $c'_k$ is the updated category center, $v_m, v_n$ are marked data and unlabeled data respectively, and $\mathbf{1}_k(\tilde{y}_n)$ is the judgment corresponding to unlabeled data. Other functions, that is, $\mathbf{1}_k(\tilde{y}_n) = \mathbf{1}[\tilde{y}_n = k, \tilde{d}_{n,k} < r_k]$.

For the test sample, the prediction method used in this paper is as follows:

$$\tilde{y}_* = \operatorname{argmax}_k P(k | v_*, \{c_k\})$$
$$= \operatorname{argmax}_k \frac{\exp(-\|v_* - c_k\|_2^2)}{\sum_{k'} \exp(-\|v_* - c_{k'}\|_2^2)} \quad (8)$$

Where: $\tilde{y}_*$ is the prediction category and $v_*$ is the test sample.

On this basis, the supervised loss based on cross-entropy can be expressed as:

$$\mathcal{L}_s = -\frac{1}{T} \sum_i \log P(y_i^* | x_i^*, \{c_k\}) \quad (9)$$

Where: $\mathcal{L}_s$ is the supervised loss, $T$ is the number of test samples, $P$ is the conditional probability, and $x_i^*$ is the original expansion feature vector.

Corresponding to the unsupervised loss, compared with the supervised loss, the unsupervised loss is more complex, for the sample points within the decision radius, the loss function structure is similar, can be written:

$$\mathcal{L}_p = -\frac{1}{M'} \sum_n \log P(\tilde{y}_n | x_n, \{c_k\}) \quad (10)$$

Where: $\mathcal{L}_p$ is the supervised loss, $M'$ is the number of unlabeled samples within the decision radius, $P$ is the conditional probability, $x_n$ is the unlabeled original expanded feature vector within the decision radius.

And for the sample points outside the decision radius, this paper uses consistency regularization to calculate the unsupervised loss:

$$\mathcal{L}_u = \Omega(x; \theta)$$
$$= \| f_{\mathrm{MLP}}(\operatorname{perb}(x); \operatorname{perb}(\theta)) - f_{\mathrm{MLP}}(x; \theta) \|_2^2 \quad (11)$$

where: $\mathcal{L}_u$ is the unsupervised loss, $\Omega$ is the consistency regularization operation, $x$ is the unlabeled original expanded feature vector outside the decision radius, perb denotes the random perturbation operation, and $\theta$ denotes the random discard rate.

To summarize, the final loss function is the sum of the three components, namely

$$\mathcal{L}_{final} = \mathcal{L}_s + \lambda \mathcal{L}_p + \mu \mathcal{L}_u \quad (12)$$

where: $\mathcal{L}_{final}$ is the final loss function and $\lambda, \mu$ is the corresponding weight coefficient.

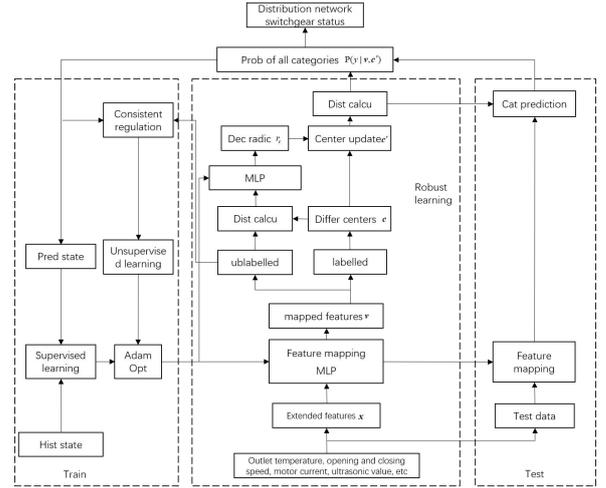

Fig. 3 Illustration of procedures of the proposed method

## 5 Calculation example analysis
### 5.1 Data source and analysis

In this paper, a set of data is collected from the operating event records of switchgear sensing terminal equipment in a certain region from May to October 21, which includes 878 records and all their corresponding feature vectors of a single switchgear on a single day. Before using the data as a program calculus, the data needs to be classified. First, the status of each switchgear for the day was evaluated according to the seven different categories described in Section 3.4, and some of the results are shown in Table 1, where the labeled sample data totaled 400 records and the unlabeled sample data counted 478 records. Secondly, the sample data are divided into 3 categories, as shown in Table 2, the labeled data are divided into two parts, half of which are used as training samples, the other part is used for example testing, and the remaining unlabeled part is used to test the effect of the weakly supervised learning strategy proposed in this paper on the improvement of the prediction accuracy.

### 5.2 Analysis of results based on the proposed semi-supervised learning framework

The original expanded feature vector data are first subjected to preprocessing operations including wavelet denoising, outlier removal, and linear normalization. In general, the mechanical characteristics of circuit breakers are usually analyzed by using cost-effective resistive angular displacement sensors combined with opening and closing coil current waveforms. In engineering applications, the operating environment and mechanical vibration of circuit breakers can reduce the accuracy of parameter calculation. The above factors can lead to an increase in the amount of deviation of the angular displacement sensor. Therefore, the calculation of derivative is used here to remove the interference point, i.e., the derivative threshold is set. If when the derivative of a point is too high, the point is judged as an interference point.

Table 1   Some switchgear working condition evaluation results

| Record number | Abnormal feature quantity | State category |
|---|---|---|
| 1 | Ultrasonic value at circuit breaker $c_1 = 18.2$ dB<br>Ultrasonic value at busbar $c_2 = 16.4$ dB<br>Ultrasonic value at line $c_3 = 17.0$ dB<br>Closing time $q_5 = 38.1$ ms<br>Opening time $q_6 = 69.2$ ms | 7 |
| 17 | Temperature of A-phase outgoing line $r_1 = 85.6\,°C$<br>Ambient temperature $t_1 = 27.2\,°C$<br>Loop bus insulation resistance $f = 281$ M$\Omega$ | 6 |
| 25 | Breaking speed $q_2 = 0.87\ m \cdot s^{-1}$<br>Breaking time $q_5 = 50.4$ ms<br>Peak current of ground cutter drive motor $l_6 = 1.2$ A | 2 |

Table 2   Distributions of different working conditions of samples

| Category | Type Number | | | | | | | Total |
|---|---|---|---|---|---|---|---|---|
| | 1 | 2 | 3 | 4 | 5 | 6 | 7 | |
| Training set | 34 | 16 | 24 | 40 | 38 | 18 | 30 | 200 |
| Test Sets | 46 | 12 | 20 | 48 | 34 | 14 | 26 | 200 |
| Not labeled | / | / | / | / | / | / | / | 478 |

The pre-processed expanded feature vector is feature mapped using a multilayer perceptron. The dimension of the input layer of the multilayer perceptron is 24, the dimension of the output layer is 8, the number of intermediate hidden layers is 2, the number of neurons contained in each layer is 48, 12, and the activation function is ReLU function. The mapped vectors are able to predict the operating state of the distribution switch, and here the center of each category is set to be the average of the feature vectors under the category. Where the representation vector of the labeled data corresponds to the category all the time, while the unknown labeled data need to predict the category center to predict the category first, and then use the feature vector to correct the center position until the model converges.

Table 3 compares the prediction accuracies under strongly supervised learning and weakly supervised learning conditions. The table shows that the average accuracy increases by 14% under weakly supervised conditions, and the performance is significantly improved in all states. Under the small sample condition, the reasonable use of unlabeled samples can mitigate the overfitting phenomenon brought by the complex model. In addition, due to the small number of samples, it is easy to be confused with other states, and further differentiation of the representation vectors by unlabeled data is needed. Therefore, the accuracy of each category, switchgear with arc extinguishing capability problems improved the most under weakly supervised learning conditions.

Table 3   Prediction results of the proposed model

| Working condition | Accuracy | | Number |
|---|---|---|---|
| | Strongly supervised | Weakly supervised | |
| 1 | 82.6% | 91.3% | 46 |
| 2 | 66.6% | 83.3% | 12 |
| 3 | 80.0% | 90.0% | 20 |
| 4 | 79.2% | 91.6% | 48 |
| 5 | 76.4% | 94.1% | 34 |
| 6 | 57.1% | 85.7% | 14 |
| 7 | 76.9% | 92.3% | 26 |
| Average | 77.0% | **91.0%** | 200 |

The correlation analysis can determine the key features under different operating conditions: 1) Normal condition: all data are within the normal range, no abnormal readings, and the equipment loading rate is within the designed operating range. 2) Operating mechanism stall: Focus on the opening and closing speeds, travel, energy storage motor, drive motor runtime and peak current. Hysteresis usually leads to changes in the speed and stroke of switching operations, as well as extended motor running time and changes in current. 3) Insulation Failure: the main concern is insulation resistance readings, which usually show a decrease. In addition, humidity readings in high humidity environments can be an important factor. 4) Empty-close phenomenon: Focus on the opening and closing speeds and strokes, as well as the ambient temperature. In a low temperature environment, the movement of the mechanism may be resisted, resulting in the switch not operating properly. (5) Mechanical failure: need to focus on the opening and closing speed, stroke, and the equipment loading rate. Mechanical failure can lead to changes in the speed and stroke of operation, but also related to changes in equipment load rate. (6) Accidental tripping: need to focus on the actual opening and closing status of the circuit breaker and ground knife, as well as the three-phase outgoing temperature. Accidental tripping will lead to changes in the state of the circuit breaker and ground knife, and will cause overheating of the line. (7) secondary equipment issues: focus on the ambient temperature and humidity, as well as equipment load factor. Secondary equipment is usually greatly affected by environmental conditions, and also related to changes in equipment load rate.

In addition, considering that the model proposed in this paper contains multiple modules, the effect of different modules on the prediction accuracy of the model is explored here using ablation experiments, and the results are shown in Table 4. It can be seen that the complete model has the highest prediction accuracy, and the feature mapping module, decision radius function, unsupervised error, and consistency regularization all contribute to the model prediction accuracy, among which the feature mapping module contributes the most, compared with the direct use of the original high-dimensional feature vectors, the module filters and adjusts the variables, so it is easier to distinguish between different categories of samples in the mapped space. The decision radius function, unsupervised error, and consistency regularization further strengthen the robustness of learning and improve the model accuracy by introducing unlabeled data.

Table 4  Results of ablation experiments

| Model | Accuracy |
| --- | --- |
| Complete model | **91.0%** |
| Without feature mapping module | 75.4% |
| Does not use decision radius to distinguish unlabeled data | 87.3% |
| Does not use unsupervised error to update model parameters | 86.4% |
| Not using consistent regularization error to update model parameters | 89.3% |

**5.3 Comparison of Common Classification Models**

This section compares the accuracy of the proposed model with the commonly used classification models (Support Vector Machine (SVM), KNearest Neighbor (KNN) and Convolutional Neural Networks (CNN)). Among them, Support Vector Machines utilize kernel functions to map input feature vectors to linearly divisible intervals, and use hyperplanes to partition feature points. The nearest neighbor algorithm makes predictions based on the nearest K neighbor feature vectors of known categories to the unknown feature vector. Convolutional neural network utilizes convolutional kernel to fuse and map the feature vectors, and inputs the processed features into the neural network for prediction.

The specific implementation details are as follows: the input features of SVM and KNN are feature vectors mapped by multilayer perceptron, where SVM uses polynomial kernel function, and its output is the switching device operating state, and the hyperparameters in KNN are determined using cross-validation, which takes the value of 10 in this paper. $K$ The input features of the one-dimensional CNN are the original feature vectors, the number of intermediate convolutional layers is 2, the size of the convolutional kernel used in each layer is 10, 6, the number of neurons in each layer including the input and output layers is 24, 48, 12, 7, respectively, and in order to alleviate the problem of overfitting, the stochastic deactivation rate [27] is set to be 0.2 and the learning rate is set to be 0.0002.

In order to reflect the robustness of the proposed method, and considering the cost of data collection and labeling, the training set is gradually reduced during the experimental process. The changes in model performance corresponding to different sizes of training sets are shown in Fig. 4. As can be seen from the figure, the method proposed in this paper is much less affected by the number of labeled data than the remaining three methods, and the test accuracy is obviously at a higher level. In summary, the method proposed in this paper is more suitable for the state diagnosis task of power distribution switchgear in real scenarios.

Table 5  Results comparison of different methods

| Working state | Model | | | | Number |
| --- | --- | --- | --- | --- | --- |
| | RLN | SVM | KNN | CNN | |
| 1 | 91.3% | 67.4% | 56.5% | 78.2% | 46 |
| 2 | 83.3% | 41.6% | 50.0% | 58.3% | 12 |
| 3 | 90.0% | 65.0% | 55.0% | 70.0% | 20 |
| 4 | 91.6% | 66.7% | 62.5% | 70.8% | 48 |
| 5 | 94.1% | 64.7% | 52.9% | 73.4% | 34 |
| 6 | 85.7% | 50.0% | 42.9% | 57.1% | 14 |
| 7 | 92.3% | 76.9% | 57.6% | 65.4% | 26 |
| Average | **91.0%** | 65.0% | 56.0% | 70.5% | 200 |

The final results are shown in Table 5, and it can be seen that the proposed method (Robust Learning Network, RLN) is much better than the remaining three models. This is due to the fact that the proposed method makes the parameter estimation more reasonable with small samples by using unlabeled samples. Secondly, the feature mapping based on multilayer perceptron and the category-centered updating strategy considering the confidence level optimize the model learning process for effective optimization, which further increases the superiority of the proposed model.

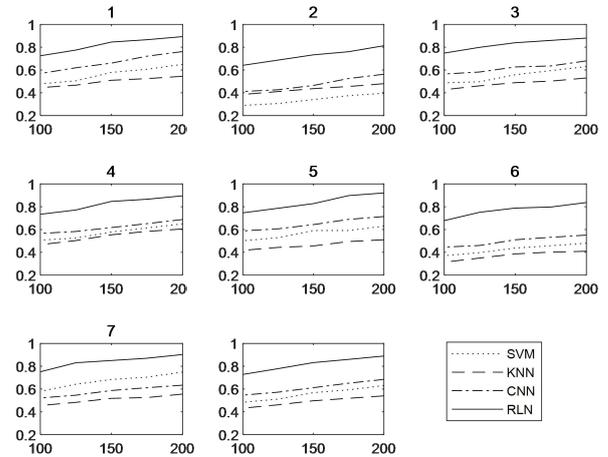

Fig.4 Relationship between prediction accuracy and training set size for different models

## 6 Conclusion

In this paper, we propose a robust learning-based state diagnosis method for distribution switchgear, which synthesizes multi-source data to provide a basis for state maintenance. The main conclusions are as follows:

(1) Construct the expanded feature vector containing environmental information, temperature information, switching position information, motor operation information, local discharge condition, insulation condition, etc., and map the original high-dimensional vectors with the help of multilayer perceptron, which effectively improves the prediction accuracy.

(2) In the feature space, the decision radius is introduced to classify the unlabeled data, and the unsupervised error and consistent regularization error are proposed to update the decision radius and the feature mapping module, which effectively utilizes the unlabeled data to achieve robust learning.

(3) Experiments based on field data show that the proposed model in this paper improves the prediction accuracy by 20.5% compared with the existing model, and by 14.0% compared with the same model under strong supervision, which proves the reasonableness of the method. It provides a worthy idea for the task of distribution network switchgear condition assessment and diagnosis.